# AUTOREGRESSIVE DESCRIPTION OF BIOLOGICAL PHENOMENA

VASILE V. MORARIU[1], CĂLIN VAMOŞ[2], ALEXANDRU POP[3], ŞTEFAN M. ŞOLTUZ[2,4],
LUIZA BUIMAGA-IARINCA[1], OANA ZAINEA[1]

[1]*National Institute of Research and Development for Isotopic and Molecular Technologies, Department of Molecular and Biomolecular Physics, 400293, Cluj-Napoca, Romania*

[2]*"T. Popoviciu" Institute of Numerical Analysis, Romanian Academy, 400110 Cluj-Napoca, Romania*

[3]*Astromomical Institute of the Romanian Academy, Astronomical Observatory, 19 Ciresilor, 400487 Cluj-Napoca, Romania*

[4]*Departamento Matematicas, Universidad de los Andes, Carrera 1 No. 18A-10, Bogota, Columbia*

Many natural phenomena can be described by power-laws of the temporal or spatial correlations. The equivalent in frequency domain is the 1/*f* spectrum. A closer look at various experimental data reveals more or less significant deviations from a 1/*f* characteristic. Such deviations are especially evident at low frequencies and less evident at high frequencies where spectra are very noisy. We exemplify such cases with four different types of phenomena offered by molecular biology (series consisting of the atomic mobility of the protein main chain), cell biophysics (flickering of red blood cells), cognitive psychology (series of mentally generated series of apparent random numbers) and astrophysics (the X-ray flux variability of a galaxy). Some of these cases can be better approximated by AR(1) – a first order autoregressive model, which is a short-range memory model – than by a 1/*f* model (long-range memory). The same spectra can be more or less easily confused and/or approximated by power-laws. On the other hand, an AR(1) model is only a zero approximation, which can be improved if more complex short-range correlation models, such as high order AR(*p*), ARMA, FARIMA models, are used. A key step to detect non-power laws in the spectra, already suggested by Mandelbrot, is to average out the spectra.

## 1. Introduction

Many processes in nature exhibit temporal or spatial correlations that can be described by power-laws. Examples can be found in physical, biological, social and psychological systems [1, 2, 3, 4, 5]. In case of a spectral analysis, the power-law is of the type $P = 1/f^{\beta}$, where f is frequency and $\beta$ is the long-range correlation exponent. Its value is $0 < \beta < 2$ for most of natural processes. A 1/*f* spectrum is diagnosed by looking at the double log plot which should be linear over the whole range of frequency scale. The slope of the linear fit is the correlation exponent $\beta$. However, a general problem encountered in the spectral analysis of various fluctuating systems is that the spectrum is quite noisy. Mandelbrot's recommendation, that apparent 1/*f* spectra should be averaged before further interpretation [4], has often been overlooked including previous work [6, 7, 8, 9]. Mandelbrot considers that non-averaged spectra lead to "unreliable and even meaningless results" [4]. An immediate consequence of the averaging procedure is that deviation from a power-law description of the spectra may be better disclosed. Sometimes, such a deviation can be obvious even if the spectrum is not averaged.

On the other hand, the non-averaged spectrum can be often reasonably fitted by a straight line, i.e. described by an apparent power-law. Such an example is illustrated in figure 1 for a heat shock protein (Protein Data Bank code: 1gme). The linear fit of the power spectrum appears to be reasonable good, yet it can be easily noticed the plateau at low frequencies (figure 1a). A simple averaging procedure can result in a clear non-1/$f$ spectrum (figure 1b).

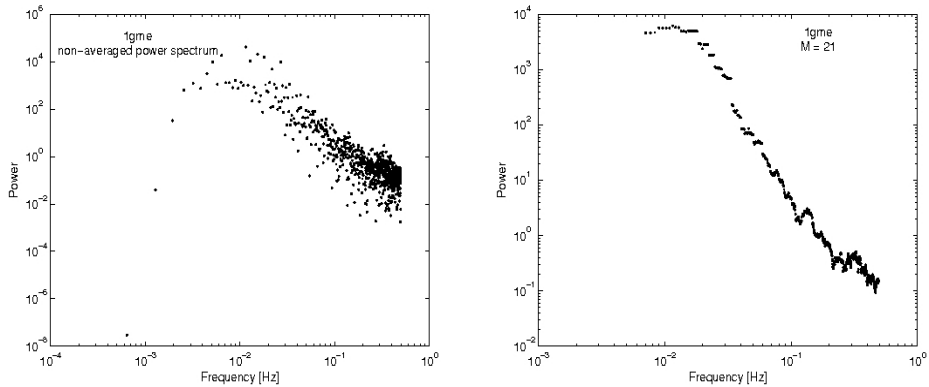

Fig. 1. a) Non-averaged power spectrum for the atomic mobility in the backbone of a heat shock protein (PDB code: 1gme), which might suggest a long-range correlation interpretation. b) The averaged power spectrum with $M$ =21 terms. It clearly indicates the short-range correlation (deviation from linearity). PDB stands for Protein Data Bank, available at http://www.rcsb.org.

However, on many occasions the non-averaged spectrum can be easily confused with a linear dependence of the double log plot. In fact, the averaging of the spectrum can indicate the deviation form linearity of the double log plot of the spectrum.

Consequently, two confusions may occur when dealing with non-averaged spectra: either a hidden non-power law remains undisclosed, and/or non-power laws are easily misinterpreted as power-laws. This work discusses the deviation from power-law behavior, which is characterized by two features: leveling of the spectrum at low frequencies, and a tendency for leveling at high frequencies, where the shape of the spectrum is blurred by the high level of noise ("the Spanish moss" as coined by [4]).

The idea of this work is twofold: first, to disclose cases of non-power laws by a simple averaging of spectra from widely different areas of science, and second, to offer an interpretation of such spectra by using an autoregressive model AR(1). This is basically different from a power-law description. While the meaning of a power-law is associated with long-range correlation or memory, the autoregressive model describes a system with short-range memory. The long-range memory is described in terms of a long-range correlation exponent $\beta$, while the short-range memory is characterized by the strength interaction $\varphi$ among consecutive terms. The literature already reported cases where astrophysical and psychological phenomena are described by autoregressive models rather than by power-laws [10, 11, 12]. Details of their approach differ to some extent.

The example from astrophysics is included in this work in order to check out our method of calculation against a different methodology.

The spectral approach in the present work was done for two reasons: first, many results in the literature are presented in spectral form and the present work was born from the observation that 1/$f$-like spectra show clear deviations from a power-low , and second, the spectral description can be easily done in an analytical manner.

This paper is organized as following: first, the main spectral features of an autoregressive model AR(1) are described for various interaction strengths among the terms of the series. Then, four classes of phenomena are described by this model: a bio-molecular process (the series of atomic mobility in the main chain of some proteins), a cell biophysics case (fluctuation of geometric parameters characterizing the flickering phenomenon of human red blood cells), a cognitive process (series of "random" integers generated by human subjects) and, finally, the procedure is checked against an astrophysical phenomenon (X-ray emission of a galaxy) where a more complex calculation based on an AR(1) model has already been published [10]. It will be shown that an identical result is obtained with a more simple procedure. All cases proved to be described by non-power laws, although they can be easily confused with 1/*f* spectra if they are not averaged. Many of these cases can be described by the AR(1) model, while others could be possibly described by higher order autoregressive models.

## 2. The spectral characteristics of the autoregressive model

A discrete stochastic process $\{X_n, n=0, \pm1, \pm2,...\}$ is called autoregressive process of order $p$, denoted AR($p$), if $\{X_n\}$ is stationary and for any $n$:

$$X_n - \varphi_1 X_{n-1} - ... - \varphi_p X_{n-p} = Z_n \tag{1}$$

where $\{Z_n\}$ is a Gaussian white noise with zero mean and variance $\sigma^2$. The real parameters $\varphi_i$, $i=1, .., p$, can be interpreted as a measure of the influence of a stochastic process term on the next term after $i$ time steps. The properties of AR($p$) processes have been studied in detail and they are the basis of the linear stochastic theory of time series [13] and [14]. Equation 1 has a unique solution if the polynomial $\Phi(z)=1-\varphi_1 z-...-\varphi_p z^p$ has no roots $z$ with $|z|=1$. If in addition $\Phi(z) \neq 1$ for all $|z| > 1$, then the process is causal, i.e. the random variable $X_n$ can be expressed only in terms of noise values at previous moments and at the same moment.

The spectral density of an AR($p$) process is:

$$f(\upsilon) = \frac{\sigma^2}{2\pi} \frac{1}{\left|\Phi(e^{-2\pi i \upsilon})\right|^2}, -0.5 < \upsilon \leq 0.5, \tag{2}$$

where $\upsilon$ is the frequency. For an AR(1) process, the spectral density in equation 2 becomes:

$$f(\upsilon) = \frac{\sigma^2}{2\pi} \frac{1}{1+\varphi^2 - 2\varphi\cos 2\pi\upsilon}, -0.5 < \upsilon \leq 0.5, \tag{3}$$

where $\varphi$ is the only parameter $\varphi_i$ in this case. The above mentioned formulas are valid for ideal stochastic processes of finite length.

The time series found in practice have a finite length and usually they are considered realizations of a finite sample of an AR(1) process of infinite length. Therefore, the changes of the equations 2 and 3 have to be analyzed for a sample with finite length $\{X_n, n=1, 2, ..., N\}$ extracted from an infinite process $\{X_n, n=0, \pm1, \pm2, ...\}$. A detailed analysis of the power spectrum of the AR(1) process and the influence of the finite length is contained in [15]. In this paper some of the main conclusions are discussed.

The sample estimator of the spectral density is the periodogram:

$$I_N(v) = |A_N(v)|^2 , \qquad (4)$$

where $A_N(v)$ is the discrete Fourier transform of the sample:

$$A_N(v) = \frac{1}{\sqrt{N}} \sum_{n-1}^{N} X_n e^{2\pi i n v} , \qquad (5)$$

Since the sample contains a finite number of components, there are only $N$ independent values of $A_N(v)$ and $I_N(v)$. Usually, these values are computed for the Fourier frequencies $v_j = j/N$, where $j$ is a integer satisfying the condition $-0.5 < v_j \leq 0.5$. The periodogram of an AR($p$) process is an unbiased estimator of the spectral density:

$$\lim_{N \to \infty} \langle I_N(v_j) \rangle = 2\pi f(v), \qquad (6)$$

where $(v_j - 0.5N) < v \leq (v_j + 0.5N)$ [13]. Hence, increasing the sample length $N$, while the time stept is kept constant, the average periodogram becomes a better approximation of the spectral density (equation 6). However, a single periodogram is not a consistent estimator, because it does not converge in probability to the spectral density, i.e. the standard deviation of $I_N(v_j)$ does not converge to zero, and two distinct values of the periodogram are uncorrelated, no matter how close the frequencies are when they are computed.

Usually, the spectral density and the periodogram are plotted on a log-log scale. The logarithmic coordinates strongly distort the shape of the graphic because by applying the logarithm, any neighborhood of the origin is transformed into an infinite length interval and the value of $f(0)$ can not be plotted. For a sample with $N$ terms, the first value of the spectral density is obtained for the minimum frequency $v_{min} = 1/N$. Figure 2a shows the spectral density in equation 3 for $N = 1024$, $\sigma = 1$ and different values of the parameter $\varphi$. One can observe that the AR(1) processes are fractal-like ones. For $\varphi = 0.90$ and especially for $\varphi = 0.99$, a significant part of the power spectrum is almost linear with a slope equal to $-2$, which corresponds to $\beta = 2$. A significant part of the spectrum could be regarded as linear for smaller value of $\varphi$ (for example $\varphi = 0.5$ in figure 2a).

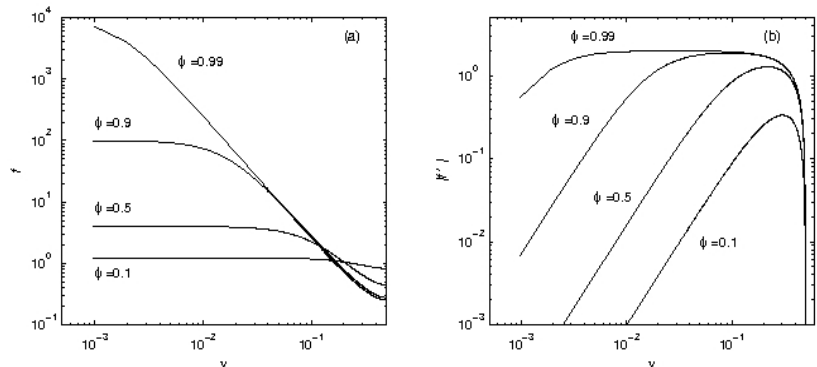

Fig. 2. The spectral density (a) and the absolute value of its derivative (b) of an AR(1) process for $N = 1024$, $\sigma = 1$ and different values of the interaction factor among close terms $\varphi$.

In order to verify this behavior of the AR(1) spectrum, figure 2b includes the derivative of the spectral density from equation 3 in log-log coordinates:

$$f'(\upsilon) = -\upsilon \frac{d}{d\upsilon}(\ln f(\upsilon)) \tag{7}$$

One can notice that for $\varphi \geq 0.9$ there is a region where $f' \cong -2$. There is only a maximum value of $f'$ for $\varphi < 0.9$ which corresponds to the center of the "linear" (or "fractal") region of the power spectrum.

For small frequencies, the AR(1) spectral density is strongly stretched in log-log coordinates such that a plateau appears (figure 2a) with a value given by:

$$f(0) = \frac{\sigma^2}{2\pi(1-\varphi)^2} \tag{8}$$

From equation 3 it follows that the plateau corresponds to the small values of $\nu$, when the variable term at the denominator can be neglected in comparison with the constant term. Using the quadratic approximation of the cosine function, the condition that the graph of the AR(1) power spectrum has a plateau $\nu < (1-\varphi)/2\pi\sqrt{\varphi}$ is obtained. If $\varphi$ tends to 1, the plateau appears at smaller values of the frequency. Therefore, if $N$ is large enough, the periodogram of an AR(1) sample has a plateau at small frequencies (if $N$ is large, then $\nu_{min} \to 0$).

A time series $\{x_n, n=1, 2, ..., N\}$ as a realization of a sample $\{X_n, n=1, 2, ..., N\}$ from an AR(1) process is considered. Applying the discrete Fourier transform (equation 4) to the time series $\{x_n\}$, and then computing the periodogram (equation 5), values randomly distributed around the spectral density (equation 3) of the AR(1) are obtained. Since the periodogram is not a consistent estimator, by increasing the length $N$ of the sample, the periodogram fluctuations around the theoretical spectral density are not reduced. Consistent estimation of the spectral density may be obtained using averaging of the periodogram on intervals with length of magnitude order of $\sqrt{N}$ [13]. Choosing the optimum weight function is a difficult task, because, if the periodogram is smoothed too much, then the bias with respect to the theoretical spectrum can become large. From various weight functions [16] the simplest one is used, i.e. the averaging with equal weights on symmetric intervals containing $M$ Fourier frequencies, with $M = 1, 3, 5, ..., 21$. Then, the averaged periodogram contains $N - M + 1$ values, because for the first and last $(M-1)/2$ values of the periodogram, the symmetric averaging can not be performed.

Let us consider that an AR(1) model for an averaged periodogram is to be found, i.e. to find the values of the parameters $\varphi$ and $\sigma$. The minimum of the quadratic norm of the difference between the averaged periodogram and the theoretical spectral density of the AR(1) model has to be determined. The sample standard deviation of the time series and $\varphi = 0$ are used as initial values for the optimization algorithm.

### 3. Data and methods

The data considered for analysis in this section belong to different areas of life sciences:

a) Molecular biology.

A protein consists of a chain of atoms built upon a backbone ($-N - C\alpha - C-$) having various lateral groups of atoms. The structure and dynamics of a protein is characterized by the atomic coordinates and the temperature factor, or the Debye-Waller factor, which are determined from the X-ray diffraction of protein crystals. The temperature factor can be regarded as a characteristic parameter for the atomic mobility. These data are available in Protein Data Bank (PDB), at http://www.rcsb.org, for a large number of proteins. We further consider the series of the temperature factors of the protein backbones, as they represent a series of data consisting of a natural succession of terms like in a time series. The organization in such structures can be characterized by looking at their correlation properties. In a previous article it has been claimed that long-range correlation is present in such series [9], and this has been further analyzed in terms of long-range correlation [10, 11, 12]. This interpretation is reconsidered in the next section. Twelve proteins were randomly selected (table 1). The *.text* file associated to each protein includes all the information obtained from X-ray crystallography (atomic coordinates, temperature factor, resolution, etc.), from which the temperature factors of the atoms in the backbone main chains were extracted (for a detailed explanation see [7]).

b) Cell biophysics – flickering of human red blood cells (RBCs).

Cell membrane undulation is a common phenomenon in the world of living cells. By far the best known is the red blood cell shape fluctuations, also known as flickering. They can easily be observed by optical microscopy. Such fluctuations in RBCs consist of sub-micron, out-of-plane displacements of the cell membrane in the frequency range of 0.3-30 Hz.

Previous studies performed investigations on RBCs adhering firmly and irreversibly to the glass substratum. In the present study free floating cells were investigated. This is an advantage as there are no mechanical restrictions imposed on cells, whereas the disadvantage is due to various motions of the cell, that induce non-stationary contributions to the basic fluctuations in the time series. Such motions either cause defocusing or an apparent change in the plan projection of various parameters. All these motions modify the geometrical parameters of the cell in an apparent manner for reasons others than the flickering itself. As a general rule, these phenomena were minimized by allowing some time for the sample to settle down and by keeping the time of image recording at a short interval. However, these phenomena may persist to various degrees and appear to be non-stationary contributions in the series of data. They can be removed by the detrending procedure mentioned below in this section.

Venous blood samples for experiments were obtained from healthy adult donors, via withdrawal into sterile vacuum tubes containing 3.8% natrium citrate. The RBCs were separated from the blood by centrifugation at 1200g for 10 min and washed three times in an isotonic saline buffer (145mM NaCl, 5mM KCl, 5mM HEPES, pH7.4). Then, the washed RBCs were resuspended in isotonic saline buffer in much more diluted suspensions. An upright optical microscope (Olympus, Model BX 51) linked to a computer, via CMOS sensor monochrome camera (PixeLINK, Model PL-A741) with a resolution of 1280x1024 pixels @27fps, was used to capture images of RBCs. For each sample, 1024 images @10fps were captured and processed by Image J, free software [17]. The cell area for each RBC has been determined by time lapse imaging. The series were further subjected to spectral analysis.

c) Cognitive psychology.

It was previously reported an analysis of series of numbers purposely produced by various human subjects as "random" series [8, 9]. In our experiments the human subjects have been asked to dictate in a random manner numbers from 0 to 9 or just 0 and 1 [9]. A general "algorithm" used by all subjects was the non-repetition of numbers, while no subject was aware that randomness involves some repetition of the numbers. The data were previously interpreted in terms of long-range correlation. However, we noticed that most of the series revealed non-power laws of the power spectra. Also, the detrended fluctua tion analysis plots (not shown) proved that long-range correlation is not valid throughout the series. The authors were not aware at that time about the non-power law description of the process and the plots were approximated by 1/*f* spectra [9]. These experiments were reconsidered and new longer series of data were produced. The spectra were averaged and they easily revealed non-power law spectra.

d) Astrophysics.

Different type of astrophysical objects display variability phenomena featured by power-law power spectra, e.g. the light curve of the 3C273 quasar. Also, the amplitude spectrum (in log-log plots) of the intermediate polar AE Aquarii shows a-1 slope [18], while the power spectrum of its radio emission time variability revealed the presence of a red noise described by a power-law. The X-ray variability of Cygbus X-1 system [19] and of Be star γ Cassiopeiae [20] is also featured by power spectra displaying 1/*f* segments. The X-ray variability of active galactic nuclei is also known to show red noise spectra which could be quite well fitted by power-laws [21]. Deviations from the simple power-law behavior are emphasized by several authors. Thus, there are mentioned cases [22, 20, 23, 24] in which different frequencies domains are featured by different slopes, or the power spectra gradually flatten toward low frequencies. Autoregressive analysis has been reported on the variability of X-ray light curves of the active Galaxy NGC5506 [10]. This analysis is compared with the present more simple approach, by using the same data extracted form Hearc Exosat ME archive for the Seyfert galaxy NGC5506.

Many series of data have non-stationary characteristics, so the application of Fourier transform to the data results in misleading spectra. A common procedure to avoid this complication is to use detrended fluctuation analysis (DFA) [25]. This results in a correlation exponent free of the correlation introduced by the trend. However, in our case it is essential to obtain the corresponding spectrum, as the shape of the spectrum gives the relevant information (either a power-law or a non-power law is operative). Consequently, an important preliminary step is to remove non-stationary characteristics in the series. We performed detrending by subtracting a polynomial fit from the original series. The problem is to determine the right polynomial fit. We performed 1 to 20 degree polynomial fits and generally found that a polynomial degree around 10 gives the most reliable result for $\varphi$ and $\sigma$. The values of $\varphi$ and $\sigma$ also depended on the averaging procedure of the spectra so that optimizing the values of $\varphi$ and $\sigma$ involved optimizing both the detrending and the averaging procedures. This will be further discussed in section 4.

The above mentioned polynomial fitting was chosen as its accuracy is comparable to the moving average method and to an automatic method for the estimation of a monotone trend. The same work also showed that polynomial fitting for a 1/*f* noise proved to have the best performance [26].

The succession of operations can be summarized as following: *i*) Detrend the series of data by subtracting various degrees of polynomial fits; *ii*) Discrete Fourier transform of the series; *iii*) Periodogram averaging using 1-21 terms; *iv*) Fit the spectrum to an AR(1) model. The resulting parameters are the interaction factor $\varphi$ and the dispersion $\sigma$. Their values depend on the degree of the polynomial fit used for the detrending procedure, and on the number of spectral terms used for averaging procedure. *v*) If the data can be de-

*Autoregressive description of biological phenomena*

scribed by an AR(1) model, choose the final values of $\varphi$ and $\sigma$, by analyzing the plot of $\varphi$ and $\sigma$ against the polynomial degree and the number of averaging terms.

## 4. Averaged spectra and fitting with AR(1) model

An averaged periodogram for mobility of backbone atoms in human hemoglobin is shown in figure 3. The non-linearity of the periodogram is obvious; therefore, it can not be described by a power-law. On the other hand, the data can be better approximated by an AR(1) model, as shown in the same figure 3.

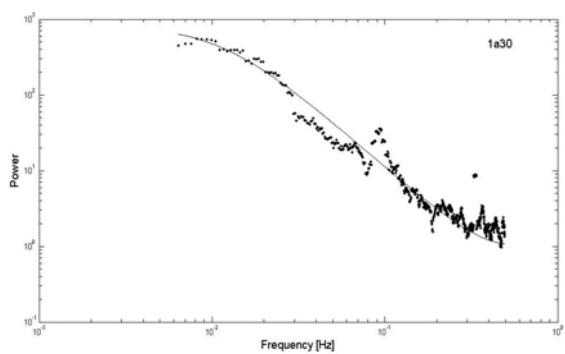

Fig. 3. Averaged periodogram and its AR(1) fit (continuous line) for the backbone atomic mobility of human hemoglobin (PDB code: 1a30). Detrending of the series was done by subtracting a 10 polynomial degree, and the number of averaged terms in the spectrum was $M=21$. The interaction factor is $\varphi=0.93$ and the noise dispersion is $\sigma=2.00$.

Further, an example of protein backbone atomic mobility for HIV-1 protease (PDB code: 4phv) is illustrated in figure 4. This example can not be described by a simple AR(1) model. The closest AR(1) model is included in figure 4 by a continuous line and it appears to be quite different from AR(1). Clearly a different model applies in this case.

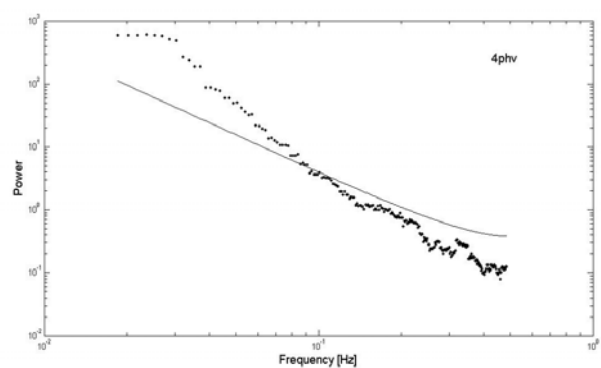

Fig. 4. Periodogram of the backbone atomic mobility for a HIV-1 protease (PDB code: 4phv). The series was preliminary detrended by subtracting a 10 polynomial degree fit. The continuous line represents an AR(1) model with $\varphi=0.99$ and $\sigma=1.13$, which was the nearest fit of the periodogram.

| No | Protein PDB code | Classification | Protein class from SCOP | AR(1) fit |
|---|---|---|---|---|
| 1 | 4phv | HIV-1 protease; (aspartic proteinase) | All beta | No |
| 2 | 1k34 | Virus/viral protein | Coiled coil | No |
| 3 | 1iiq | HIV-1 protease; hydrolase inhibitor | All beta | Yes; $\varphi = 0.927$ |
| 4 | 1gbu | Deoxy human hemoglobin | All alpha | No |
| 5 | 1fbd | Hydrolase (phosphoric monoester) | Alpha and beta | Yes; $\varphi = 0.981$ |
| 6 | 1c9m | Hydrolase | Alpha and beta | No |
| 7 | 1c9n | Hydrolase | Alpha and beta | No |
| 8 | 1a8g | HIV-1 protease complex (acid proteinase/inhibitor) | All beta | Yes; $\varphi = 0.965$ |
| 9 | 1a30 | HIV-1 protease complex (aspartic protease/inhibitor) | All beta | Yes; $\varphi = 0.923$ |
| 10 | 1qbs | HIV-1 protease; aspartyl protease | All beta | Yes; $\varphi = 0.964$ |
| 11 | 1gme | Eukaryotic small heat shock protein; chaperone | All beta | No |
| 12 | 1atr | Chaperone protein | Alpha and beta | No |

Table 1: Classification and classes of proteins investigated by the AR(1) model. All the cases with no AR(1) fit are described by higher order AR models. The results for φ were obtained for detrended series by using 10 degree polynomials and M =21 terms for averaging.

Twelve randomly selected proteins, belonging to various protein classes, were analyzed in order to find out to what extend their spectra can be described by an autoregressive AR(1) model (table 1). Table 1 also includes the information about the autoregressive model in terms of "yes" if there is a fit to our data by AR(1) and the corresponding correlation coefficient $\varphi$ is indicated or "no" if there is no such a fit.

It is found that all cases do not have long-range characteristics, i.e. the double log plot of the averaged spectra deviates significantly form linearity. About 40% of the cases can be described by an AR(1) model, while the rest of them can be described by higher order autoregressive models (table 1).

The next case is taken from red blood cell flickering. An example of a spectral analysis of the area fluctuation is shown in figure 5. The spectrum remains noisy even after the averaging method, but it can be clearly seen that an AR(1) model describes the data better than a power-law.

*Autoregressive description of biological phenomena*

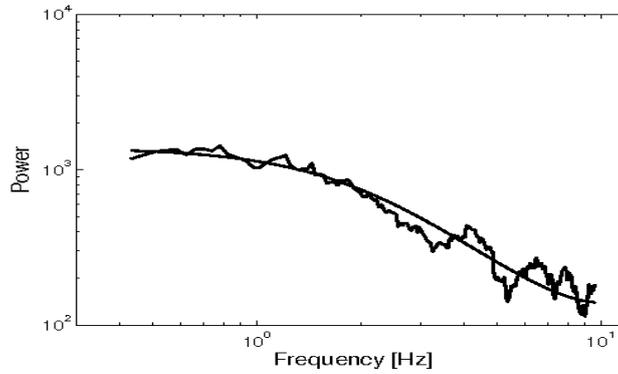

Fig. 5. Averaged periodogram for $M$ =21 terms of human RBCs area fluctuations suspended in buffer solution and its fitting by an AR(1) model (continuous line) with $\varphi$ =0.52 and $\sigma$ =17.92. The initial series was detrended by a 10 polynomial degree.

The power spectrum also indicates some oscillations around the AR(1) curve in the higher frequency range. Similar to the case of proteins, it is suitable to use a 10 degree polynomial degree for detrending and 21 terms for the averaging procedure in order to obtain a reliable value for the interaction factor $\varphi$. This choice ensures the most constant value of $\varphi$ for various averaging terms $M$. All series show that cell area fluctuation is characterized by an interaction factor $\varphi$ around $0.5 - 0.7$. The data seem to be noisier than the protein examples, yet they can be definitely classified as short-range memory systems. Further, an example from a cognitive psychology experiment is given. This case can be well modeled by an AR(1) model with a negative $\varphi$ (figure 6). From sixteen series generated by different subjects only four series could be fitted by the AR(1) model. Other series produced typical spectra of the type shown in figure 7, which can not be described by an AR(1) model.

Although this article is limited to an illustration with AR(1) model of various classes of phenomena, an example of higher order autoregressive model (figure 8) is included, which resembles qualitatively the data in figure 7.

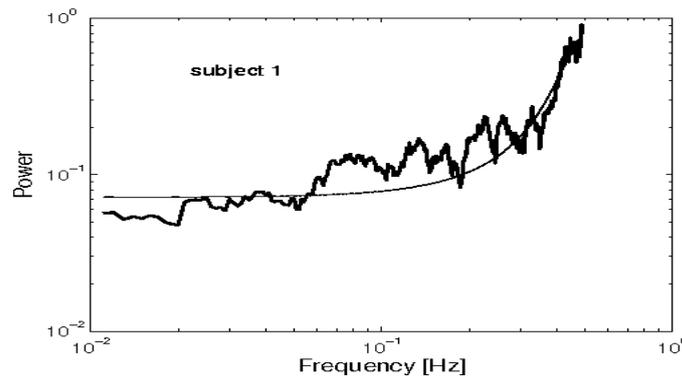

Fig. 6. Periodogram of mental series generated by the first human subject fitted by an AR(1) model (continuous line) with $\varphi = -0.53$ and $\sigma$ =0.41.Note the anti-correlation character of the spectrum.

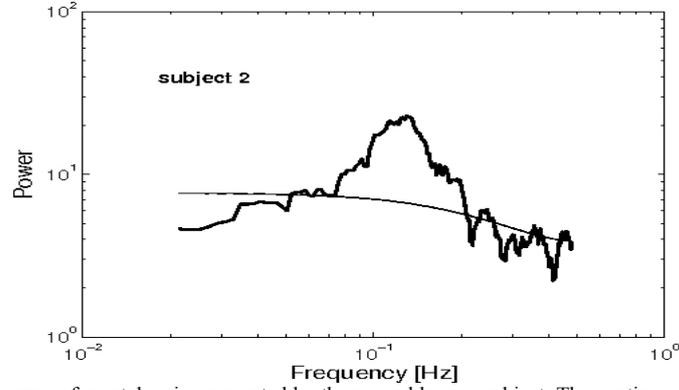

Fig. 7. Periodogram of mental series generated by the second human subject. The continuous line is the AR(1) fit, that gives $\varphi$ =0.17 and $\sigma$ =2.22. This plot illustrates that the mental series can not be described by an AR(1) model; higher order AR models should be used.

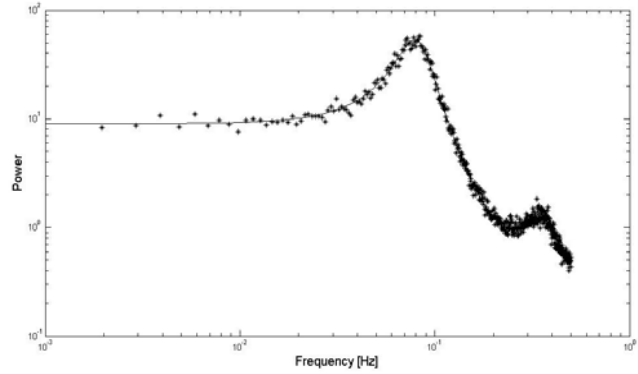

Fig. 8. Periodogram of an AR(2) model series. This AR model simulation is quantitatively similar to the mental series in figure 7 (it is characterized by a higher order AR), with $\varphi_1$ =0.8 and $\varphi_4$ =−0.3.

This might suggests that the series are described by a strong interaction among neighbors and by a weaker anti-interaction at four terms distance. In other words, the numbers produced by a subject are strongly related to each other and it is quite the opposite at longer distances. In fact, all series which were not fitted by AR(1) can be described by similar higher order AR models. These results show that mental series are generated according to different "algorithms" depending on the subject and therefore, this method of analysis may represent a promising tool of investigation for such cognitive experiments.

## 5. Checking the AR(1) fitting procedure against an astrophysical example

Our calculations have as a final result the value of $\varphi$ and $\sigma$ of the AR(1) model fitted to a particular series of data. The procedure was checked against a more complex system formed by series of astrophysical data. The variability of X-ray flux from galaxies has been previously described as flickering or 1/*f* fluctuation [10]. A specific problem in astronomical observations is the observational noise, as well as other misleading systematic effects occurring in power spectra. As a result, a specific model was considered in order to generalize AR processes and to estimate the hidden autoregressive process [10]. The model is known as the Linear State Space Model (LSSM) in which the observational noise is explicitly modeled. The results reported by König and Timmer for the first order process AR(1) are the parameter $\varphi = 0.994$ and the standard deviation $\sigma = 0.722$.

König and Timmer analysis was compared with the present more simple approach and proved to be in a very good agreement. The data series are represented in figure 9a. The time series consist of $N = 6948$ values, with the sample interval $dt = 60s$. There are only 20 values which are not equidistant. The maximum value of the sampling interval is 3310s. However, the time series are considered as being almost equidistant and further precautions will be reduced to minimum.

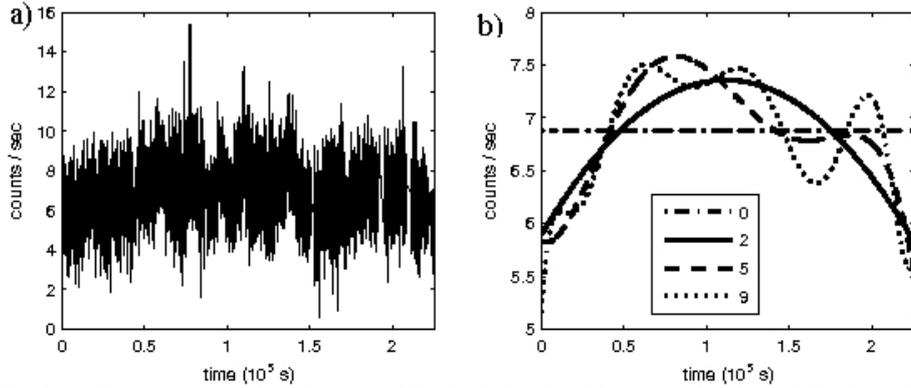

Fig. 9. a) The X-ray time series of galaxy NGC5506; b) Trends of the time series obtained for three different polynomial fittings (0, 2, 5 and 9). The trend reduces to a constant when $q = 0$. for $q = 2$, the polynomial trend describes the global shape of the signal. For $q = 5$, the polynomial trend describes the different behavior of the first and last half of the signal. The polynomial trend can follow the details of the signal when $q = 9$.

The first problem is to remove the deterministic trend that can be found by examining the shape of the signal. Timmer and König did not extract any deterministic trend in their analysis. Three different trends can be obtained by different polynomial fitting degrees (figure 9b). The degree of these polynomials equals the minimum degree of a class of polynomial trends $q$, which has very similar shapes. When the degree of the polynomial trend increases, the shape of the trend does not change monotonically. At certain polynomial degrees, the trend has more significant changes, while for the following polynomial degrees the shape remains practically unchanged. The trend reduces to a constant, which equals the average of the temporal series when $q = 0$. There is no significant improvement for a linear trend ($q = 1$), while for $q = 2$ the polynomial trend describes the global shape of the signal. Only for $q = 5$ the polynomial trend describes the different behavior of the first and last half of the signal. The polynomial trend can follow better the details of the signal when $q = 9$, however, numerical oscillations arise at the end, which can not be associated with real variations. These numerical oscillations increase with the degree of the polynomial, therefore they were not considered. The choice of these four representative values for the degree of the polynomial trend will be quantitatively confirmed by the variation of parameters characterizing the autoregressive model.

The calculation of the periodogram can not be done using equation 5, since there are missing data in twenty regions. The other values of the series are separated by time interval $dt = 30s$. Therefore, the series can be considered as equidistant and they have $T = 7532$ values, with 584 missing values. A zero value is assigned to these data, so they do not contribute in equation 5 to the value of the periodogram. The periodogram is presented in figure 10a for the signal where only the mean value ($q = 0$) was subtracted, and in figure 10b for averaged periodogram with $M = 21$ values. It can be seen that the averaging procedure results in loosing data for the lower frequencies which describe the plateau of an autoregressive model.

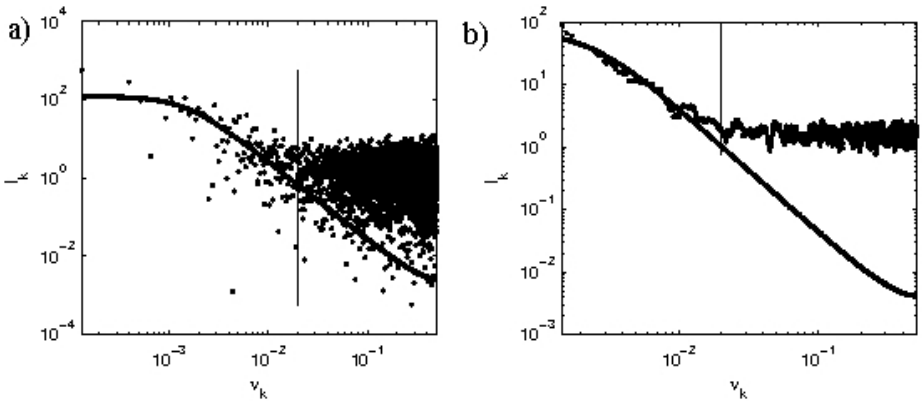

Fig. 10. a) Periodogram of the time series for galaxy NGC5506. The mean value $q = 0$ was subtracted; b) The averaged periodogram using a rectangular window with $M = 21$ values. The averaging procedure results in loosing data for the low frequencies that describe the plateau of an autoregressive model. At frequencies higher than the cutting frequency $v_0 = 0.02$, the shape of the periodogram changes to a white noise spectrum. Only one part of the spectrum, for $v > v_0$, can be modeled with an AR(1) process.

At frequencies higher than the cutting frequency $v_0 = 0.02$, the shape of the periodogram changes to a white noise spectrum, as it can be seen in figure 10b. Then, only one part of the spectrum, for $v > v_0$, can be modeled with an AR(1) process. The values $\varphi = 0.991$, $\sigma = 0.757$ and $\varphi = 0.985$, $\sigma = 0.744$ are obtained by fitting the lower frequencies part of the periodogram to an AR(1) process for the non-averaged and the averaged periodogram respectively. These values are very close to those reported by Timmer and König.

As in the former example, the averaging interval of periodogram has a strong influence on the parameters of the AR(1) model. This dependence is shown in figure 11 as a function of $M$ for different degrees of the polynomial trend. For small values of $M$, the values of $\varphi$ and $\sigma$ for different polynomial trends are quite different. Their variability for $M > 9$ is significantly reduced, therefore the averaging procedure can eliminate the fluctuation of the periodogram. This is why $M = 11$ is used in this investigation.

*Autoregressive description of biological phenomena*

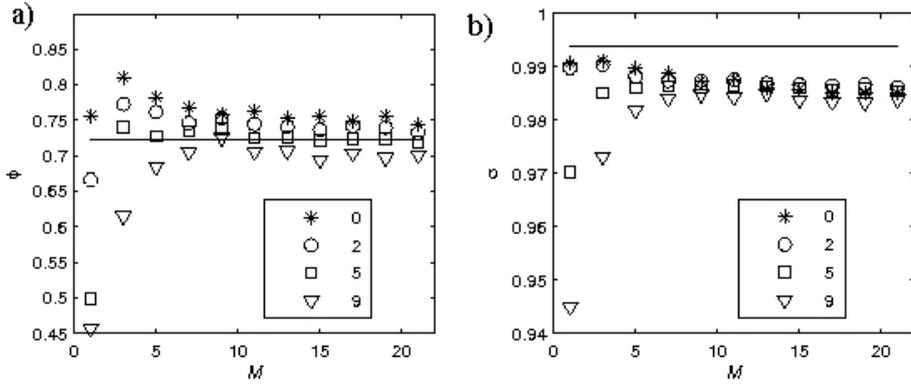

Fig. 11. a) Dependence of the parameters (a) $\varphi$ and (b) $\sigma$ for the AR(1) model on the polynomial trend degree (from 0 to 9) for different $M$ values. Continuous lines represent the values reported by [10]. For small values of $M$, the values of $\varphi$ and $\sigma$ for different polynomial trends are quite different. For $M > 9$, their variability is significantly reduced, so the averaging procedure can eliminate the periodogram of fluctuation.

Another parameter which has to be analyzed in order to fit an autoregressive model is the cutting frequency $v_0$. It can be noticed that at frequencies smaller than 0.02 the periodogram presents oscillations, which can be caused by the white noise at higher frequencies. The dependence of the autoregressive parameters on the value of $v_0$ is presented in figure 12. It shows that dispersion of the noise does not depend significantly on the periodogram interval used for the fitting procedure. However, for all the degrees of the polynomial trend eliminated from the initial signal, the value of $\varphi$ decreases with the increase of $v_0$. Only the first two values are almost equal. Such a decrease can be explained by the influence of the observational white noise in the modeled part of the periodogram. Consequently, the cutting frequency was established at $v_0 = 0.005$.

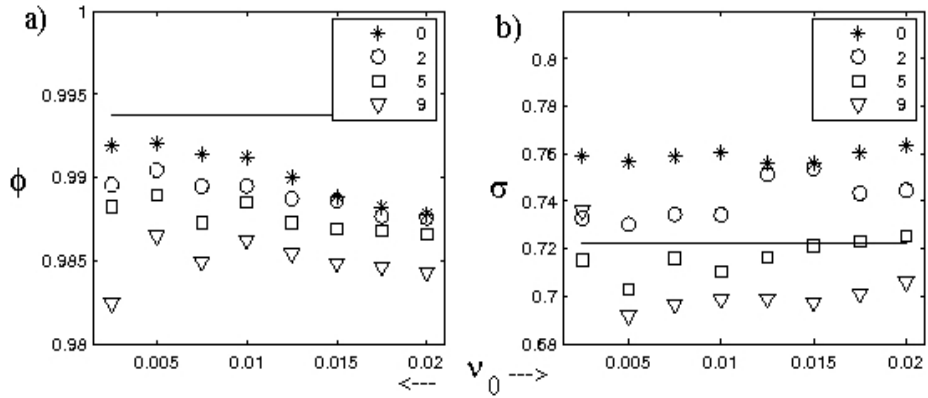

Fig. 12. The dependence of the autoregressive parameters (a) $\varphi$ and (b) $\sigma$ on the cutting frequency $v_0$.

All these results show a significant dependence of the autoregressive parameters on the degree of the polynomial trend which was eliminated from the initial signal. The dependence of the autoregressive parameters on the degree of polynomial trend for several values of the cutting frequency is presented in figure 13.

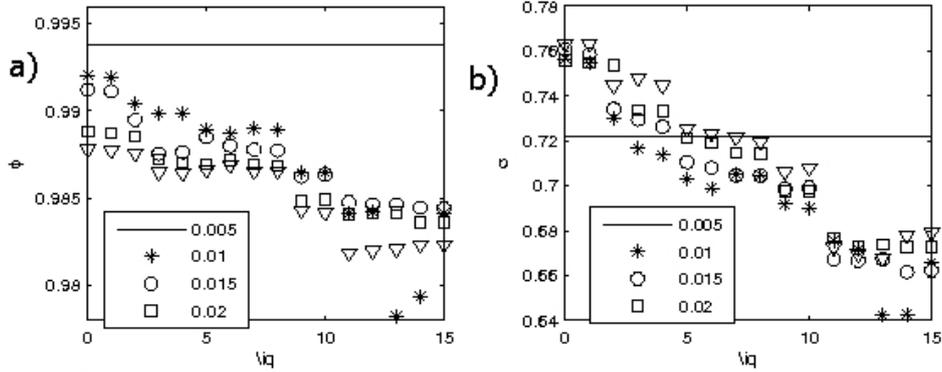

Fig. 13. The dependence of the autoregressive parameters (a) $\varphi$ and (b) $\sigma$ on the degree of polynomial trend $q$.

First, the stepwise variation at $q = 2, 5$ and $9$ can be noticed, which corresponds to the discussion mentioned at the beginning of this section. Another discontinuity at $q =11$ was no longer used, as the oscillations at the end of the interval become too large. Also, it can be noticed that the autoregressive parameters decrease as the degree of the polynomials increases. This is due to the fact that higher polynomial degrees can better describe the oscillations of the signal, while for lower degrees they are modeled by the autoregressive process. As Koning and Timmer did not remove any deterministic trend from the signal, a comparison with their results should consider the results for $q =0$. The relative error for $\varphi$ is 0.2% and 5% for $\sigma$, which confirms the spectral autoregressive modeling method of the time series proposed in this work.

## 6. Conclusions

Four different kind of fluctuating phenomena originating from molecular biology, cell biology, cognitive psychology and astrophysics proved to be short-range memory systems, and a significant percentage of them could be described by the most simple autoregressive model of first order AR(1). The characteristic parameters for this model were $\varphi$ – the strength of interaction among consecutive terms – and the dispersion $\sigma$ of the data. Phenomena can be broadly classified into three categories according to $\varphi$: a) strong interacting systems with $\varphi =0.80- 0.99$; b) medium interacting systems with $\varphi =0.5- 0.7$, and c) weak interacting systems with $\varphi =0.2- 0.4$. This classification is neither strictly delimited nor attributed to specific criteria apart from arbitrary numerical appreciation. The first category includes the X-ray variability of galaxies and protein structures. In the second category falls the flickering of blood cells and in the third category, some data for a cognitive process, such as the random generation of numbers, and DNA structure (not shown). The proportion of AR(1) models in each category may vary to some extend. However, around 1/4 to 1/3 of the cases is described by AR(1). It is also evident that higher degree autoregressive models are operative in other cases.

The parameter $\varphi$ proved to be sensitive and therefore can be profitably exploited to investigate various effects on the fluctuating system. On the other hand, such processes can easily be confused and/or approximated by power-laws. The most important step in disclosing the nature of fluctuations is to average out their spectra. Apparent 1/$f$ spectra should be cautiously treated and averaging should be compulsory. It is suspected that other phenomena are suitable for an autoregressive description, including higher order autoregressive models.

**Acknowledgements**

This work was funded by Romanian Academy for Scientific Research, Grant no. 2CEEX-06-II-96.